\documentclass[aip, apl,twocolumn,reprint,superscriptaddress,showkeys,floatfix,amsmath,amssymb]{revtex4-2}

\usepackage[utf8]{inputenc}
\usepackage[T1]{fontenc}
\usepackage{mathptmx}
\usepackage{etoolbox}
\usepackage{graphicx} 
\usepackage{dcolumn}  
\usepackage{bm}       
\usepackage{color,mathtools,bbm} 
\usepackage{wasysym}    
\usepackage{hyperref}  
\usepackage{lineno}
\usepackage{longtable}
\usepackage{footnotehyper}
\makesavenoteenv{longtable}
\usepackage{array}
\usepackage{mciteplus}
\usepackage{hypernat}
\usepackage{float} 
\usepackage{wrapfig} 
\usepackage{adjustbox}
\usepackage{braket}
\usepackage{todonotes}

\providecommand{\keywords}[1]{\textbf{\textit{Keywords:}} #1}

\begin{document}

\title{Microscopic Insights to the Ultralow Thermal Conductivity of Monolayer 1T-SnTe$_\textbf{2}$}

\author{Kemal Aziz}
\affiliation{Department of Physics, Lehigh University, Bethlehem, PA 18015, USA}
\affiliation{Department of Physics \& Astronomy, Center for Materials Theory, Rutgers University, Piscataway, NJ 08854, USA}

\author{John E. Ekpe}
\affiliation{Department of Physics, Alex Ekwueme Federal University Ndufu-Alike, Ebonyi State 482131, Nigeria}

\author{Augustine O. Okekeoma}
\affiliation{Department of Physics \& Astronomy, University of Nigeria, Nsukka, Enugu State 410001, Nigeria}

\author{Stanley O. Ebuwa}
\affiliation{Department of Physics, University of Benin, Benin City, Edo State 300283, Nigeria}

\author{Sylvester M. Mbam}
\affiliation{Department of Physics \& Astronomy, University of Nigeria, Nsukka, Enugu State 410001, Nigeria}

\author{Shedrack Ani}
\affiliation{Department of Physics \& Astronomy, University of Nigeria, Nsukka, Enugu State 410001, Nigeria}

\author{Malachy N. Asogwa}
\affiliation{Department of Physics \& Astronomy, University of Nigeria, Nsukka, Enugu State 410001, Nigeria}

\author{Richard A. Mangluhut}
\affiliation{Department of Physics \& Astronomy, University of Nigeria, Nsukka, Enugu State 410001, Nigeria}

\author{Anthony C. Iloanya}
\affiliation{Department of Physics, Lehigh University, Bethlehem, PA 18015, USA}

\author{Fabian I. Ezema}
\email{fabian.ezema@unn.edu.ng}
\affiliation{Department of Physics \& Astronomy, University of Nigeria, Nsukka, Enugu State 410001, Nigeria}

\author{Chinedu E. Ekuma}
\email{cekuma1@gmail.com}
\affiliation{Department of Physics, Lehigh University, Bethlehem, PA 18015, USA}

\date{\today}

\begin{abstract}
Two-dimensional (2D) metallic systems with intrinsically low lattice thermal conductivity are rare, yet they are of great interest for next-generation energy and electronic technologies. Here, we present a comprehensive first-principles investigation of monolayer tin telluride (SnTe$_2$) in its 1T (CdI$_2$-type, $P\bar{3}m1$) structure. Our calculations establish its energetic and dynamical stability, confirmed by large cohesive (10.9 eV/atom) and formation ($-$4.06 eV/atom) energies and a phonon spectrum free of imaginary modes. The electronic band structure reveals metallicity arising from strong Sn--Te $p$ orbital hybridization. Most importantly, phonon dispersion analysis uncovers a microscopic origin for the ultralow lattice thermal conductivity: the heavy mass of Te atoms, weak Sn--Te bonding, and flat acoustic branches that yield exceptionally low and anisotropic group velocities ($\sim 5.0\times10^3$ m/s), together with the absence of a phonon bandgap that enhances Umklapp scattering. These features converge to suppress phonon-mediated heat transport. Complementary calculations of the optical dielectric response and joint density of states reveal pronounced interband transitions and a plasmonic resonance near 4.84 eV, suggesting additional optoelectronic opportunities. These findings establish monolayer SnTe$_2$ as a 2D material whose vibrational softness naturally enforces ultralow lattice thermal conductivity, underscoring its potential for thermoelectric applications.
\end{abstract}


\maketitle

\section{Introduction}
Two-dimensional (2D) materials have attracted significant attention owing to their diverse structural, electronic, and thermal properties that arise from quantum confinement and reduced dimensionality.~\cite{zhou2025illuminating,ramalingam2020quantum,bachu2024quantum} Among these, transition metal dichalcogenides (TMDs) constitute a particularly important class of layered compounds with the general formula $MX_2$, where $M$ denotes a transition or post-transition metal and $X$ represents a chalcogen (S, Se, or Te).~\cite{lasek2021synthesis} The characteristic layered nature of TMDs arises from $X$–$M$–$X$ sandwich units stacked along the crystallographic $c$-axis, held together by weak van der Waals interactions, which facilitate exfoliation into atomically thin monolayers.

TMDs can crystallize in different polymorphs, the most prominent being the 2H (hexagonal, trigonal prismatic coordination; $D_{3h}$ symmetry) and the 1T (trigonal, octahedral coordination; $D_{3d}$ symmetry) phases.~\cite{sim2022recent,EKUMA2019383} The 2H phase is typically semiconducting and exemplified by materials such as MoS$_2$, whereas the 1T phase is generally metallic.~\cite{marinov2023ex,ulian2023structural,capobianco2022terahertz,fang2018structure,lin2014} Structural phase stability and the resulting electronic properties are dictated by orbital hybridization and coordination chemistry. Tin-based dichalcogenides, including SnS$_2$ and SnSe$_2$, are known to crystallize in the CdI$_2$-type 1T structure, in which each Sn atom is octahedrally coordinated by six chalcogen atoms.~\cite{kim2024phase} In this arrangement, a single Sn layer is sandwiched between two chalcogen layers, and adjacent monolayers interact primarily through van der Waals forces. SnTe$_2$ is anticipated to adopt a similar 1T structure (space group $P\bar{3}m1$), consistent with its post-transition metal character and the dominance of $s$- and $p$-orbitals near the Fermi level rather than $d$-orbitals, as is typical for transition-metal-based TMDs.~\cite{antonelli2022orbital,yilmaz2024evolution,aldaoseri2024,littlewood2010band,aldaoseri2024metals}

The electronic structure of 1T TMDs strongly influences their transport properties. In contrast to 2H semiconductors, 1T phases often display metallic conduction due to the higher coordination environment. This metallicity, combined with structural flexibility, makes certain 2D metals promising candidates for next-generation nanoelectronic and optoelectronic devices.~\cite{huang2014tin, kazemi2022theoretical,sokolikova2020direct} In particular, monolayer SnTe$_2$ is predicted to behave as a 2D metal with appreciable electrical conductivity.~\cite{wen2015electronic,linh2023prediction,wang2019ultralow} Understanding its fundamental physical properties, especially those governing heat and charge transport, is therefore crucial for evaluating its potential in functional applications.

Thermal properties of 2D materials are intimately tied to their phonon dispersions.~\cite{kim2015} The slope of the acoustic branches near the Brillouin zone center, given by the group velocity $v_{g} = d\omega/dk$, plays a central role in determining the lattice thermal conductivity $\kappa_{L}$.~\cite{li2024first, chen2018rationalizing} Low group velocities, combined with heavy constituent atoms and weak interatomic bonding, typically lead to suppressed $\kappa_{L}$. In the case of SnTe$_2$, the heavy mass of Te atoms and relatively soft Sn--Te bonds are expected to produce low-frequency acoustic phonon modes, strong anharmonicity, and enhanced phonon--phonon scattering (including Umklapp processes), all of which act to reduce $\kappa_{L}$. Similar trends have been reported for related systems such as SnS$_2$ and SnSe$_2$, which exhibit ultralow lattice thermal conductivities in their monolayer forms.~\cite{gu2016phonon,wang2019ultralow} The thermoelectric performance is commonly evaluated by the dimensionless figure of merit, $ZT = \tfrac{S^2 \sigma T}{\kappa_{e} + \kappa_{L}}$, where $S$ is the Seebeck coefficient, $\sigma$ the electrical conductivity, $T$ the absolute temperature, and $\kappa_{e}$ and $\kappa_{L}$ are the electronic and lattice contributions to thermal conductivity, respectively.~\cite{PhysRevB.85.085205,ZHANG2020147387} A high $ZT$ requires both a large power factor ($S^2 \sigma$) and minimal total thermal conductivity. The Wiedemann--Franz relation, $\kappa_{e}/\sigma = L T$ (with $L$ as the Lorenz number), highlights the intrinsic coupling between $\sigma$ and $\kappa_{e}$ in metallic systems, making reduction of $\kappa_{L}$ the most effective strategy to enhance $ZT$.~\cite{PhysRevB.85.085205,Yadav2019} 

Low-dimensional materials offer distinct advantages for thermoelectrics. Reduced dimensionality enhances quantum confinement of charge carriers and introduces additional scattering mechanisms that suppress phonon transport without substantially degrading electronic conduction.~\cite{Tripathi2010,Tai2008,zhou2019high,zhou2022anomalous} As a result, 2D chalcogenides and nanostructures often exhibit higher $ZT$ values compared to their bulk counterparts.~\cite{pandit2021thermal} Recent theoretical predictions indicate that monolayer SnTe$_2$, along with its structural analog SiTe$_2$, could achieve lattice thermal conductivities as low as $\sim 1.6$ W/mK at 300 K and reach $ZT$ values up to $\sim 0.7$ at elevated temperatures (900 K) under optimized conditions.~\cite{wang2012calypso} These findings underscore the promise of SnTe$_2$ as a thermoelectric material.

Motivated by these considerations, we present a comprehensive first-principles investigation of the structural, electronic, and vibrational properties of monolayer 1T-SnTe$_2$. Our central aim is to provide a microscopic explanation for the origin of its intrinsically ultralow lattice thermal conductivity. We demonstrate that this behavior arises from the combined effects of the heavy mass of Te atoms, weak Sn--Te bonding interactions, and the resulting flat acoustic phonon dispersions that give rise to exceptionally low group velocities. These factors, together with enhanced phonon–phonon scattering channels promoted by the absence of a phonon bandgap, converge to suppress heat conduction and rationalize the anomalously low $\kappa_L$ of monolayer SnTe$_2$. To establish this microscopic picture, we systematically examine: (1) the structural and energetic stability of the 1T phase; (2) the electronic band structure and orbital contributions to metallicity; (3) phonon dispersion and vibrational mode analysis as direct signatures of lattice softness; and (4) the anisotropic acoustic phonon velocities that quantitatively explain the suppression of thermal transport. In addition, we present the optical dielectric response and joint density of states, which, while secondary to the main focus, provide complementary insight into the electronic excitations and potential optoelectronic functionalities of SnTe$_2$. Collectively, these results identify monolayer SnTe$_2$ as a dynamically stable metallic dichalcogenide with exceptionally weak phonon-mediated heat transport, highlighting its significance in the broader context of energy and electronic materials research.

\section{Method}
The ground state crystal structure of monolayer SnTe$_2$ was predicted using the ab initio evolutionary algorithm as implemented in the CALYPSO software package.~\cite{wang2012calypso} Subsequent structural optimization and electronic property calculations were performed using density functional theory (DFT) within the Quantum Espresso simulation suite.~\cite{giannozzi2009quantum} The Projector Augmented Wave (PAW) method~\cite{blochl1994projector} was used to describe the electron-ion interactions, and the exchange-correlation functional was treated with the Generalized Gradient Approximation (GGA) as parameterized by Perdew, Burke, and Ernzerhof (PBE).~\cite{perdew1996generalized} The valence electron configurations considered were $Sn(4d^{10}5s^25p^2)$ and $Te(5s^25p^4)$. A plane-wave kinetic energy cutoff was set to 600 eV. The Brillouin Zone (BZ) was sampled using a $\Gamma$-centered $10\times 10\times 1$ k-point mesh for self-consistent field calculations. The convergence criteria for total energy and interatomic forces were set to $10^{-8}$ eV and $10^{-4}$ eV/\AA, respectively, with a David diagonalization threshold.~\cite{giannozzi2009quantum} To model an isolated monolayer, a vacuum spacing of 28 \AA was introduced in the direction perpendicular to the atomic plane following structural optimization. Interlayer van der Waals interactions were accounted for using the Tkatchenko-Scheffler dispersion correction scheme.~\cite{tkatchenko2009accurate} An alternative treatment using Grimme's DFT-D2 method was also employed for comparison.~\cite{Grimme2006} Structural relaxation was performed using the Broyden-Fletcher-Goldfarb-Shannon (BFGS) algorithm for variable cell optimization.~\cite{broyden1970convergence,fletcher1970new,shanno1970conditioning,goldfarb1970family} The vibrational properties, including phonon dispersion and spectroscopic intensities, were calculated using Density Functional Perturbation Theory (DFPT) as implemented in Quantum Espresso. A $2\times 2\times 1$ supercell was used for the calculation of the dynamical matrix and interatomic force constants. The optical properties were determined by calculating the frequency-dependent dielectric function within the Random Phase Approximation (RPA).

\section{Results and Discussion}

\begin{figure}[htb!]
    \centering
    \includegraphics[trim = 0mm 0mm 0mm 0mm,width=\linewidth,clip=true]{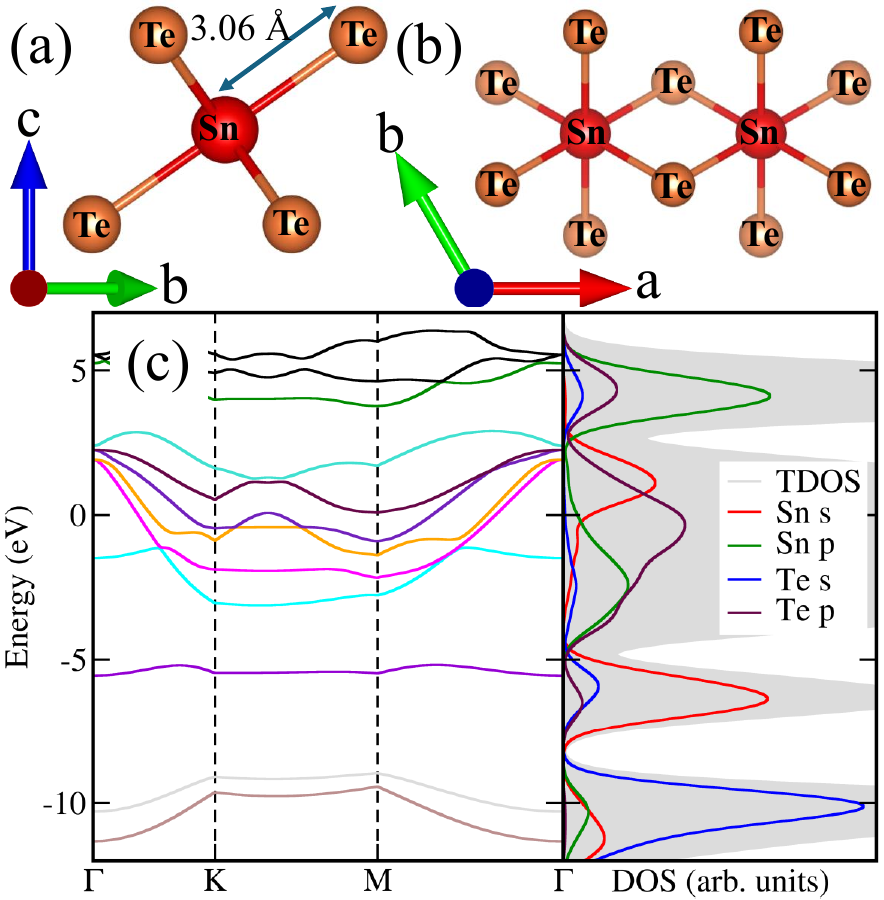}
    \caption{(a) Side and (b) top views of the crystal structure of monolayer SnTe$_2$ in the P3m1 space group. (c) Electronic band structure and projected density of states (PDOS) of monolayer SnTe$_2$. The overlap of conduction and valence bands at the Fermi level ($E_F = 0.52$ eV) confirms its metallic character. The PDOS highlights the dominant contributions of Sn and Te atomic orbitals near $E_F$, indicating hybridization between cation and anion states.}
    \label{Fig1}
\end{figure}

We present the structural, electronic, vibrational, thermal, and optical properties of monolayer SnTe$_2$ in its 1T phase (see Figure~\ref{Fig1}a \& b). The results are organized to progressively establish the stability of the material, characterize its metallic nature, and explore the lattice dynamics and phonon-mediated transport properties, followed by an analysis of the optical response. Beginning with energetic and dynamical stability, we then discuss the electronic structure, phonon spectrum, and lattice vibrations, and finally examine thermal transport and dielectric behavior. This flow highlights the interplay between bonding, conduction, vibrational motion, and optical activity, providing a comprehensive picture of SnTe$_2$ as a potential 2D material for thermoelectric and optoelectronic applications.

\subsection{Structural and Energetic Stability}
The phase stability of monolayer SnTe$_2$ was first examined via its cohesive energy $E_{\text{coh}}$, which describes the interaction of atoms with their neighbors in the 2D network and controls the mechanical stability of the whole lattice.~\cite{D1NH00113B} We define the cohesive energy per formula unit as $E_{\text{coh}} = E[\text{Sn}] + 2E[\text{Te}] - E[\text{SnTe}_2]$, where $E[\text{Sn}]$ and $E[\text{Te}]$ are the total energies of isolated Sn and Te atoms, and $E[\text{SnTe}_2]$ is the total energy of one SnTe$_2$ formula unit in the monolayer. We find $E_{\text{coh}} \approx 10.9$~eV per atom, indicating that the SnTe$_2$ monolayer is strongly bound and energetically stable. For comparison, reported cohesive energies of several transition-metal telluride monolayers in the 1$T$ structure, typically span $\sim 12$--$16$~eV per formula unit, depending on the compound and computational details.~\cite{ataca2012stable} The magnitude of the $E_{\text{coh}}$ of SnTe$_2$ highlights the strength of the Sn--Te bonds despite the absence of transition-metal $d$ orbital participation. A more stringent measure of stability is the formation energy with respect to the elemental solids. Using the cohesive energies of bulk Sn and Te (denoted $E_c[\text{Sn}]$ and $E_c[\text{Te}]$), the formation energy per atom can be written as  $E_{\text{form}} = ({E_c[\text{SnTe}_2] - E_c[\text{Sn}] - 2E_c[\text{Te}]})/{3}$. We obtain $E_{\text{form}} \approx -4.06$~eV per atom, confirming that monolayer SnTe$_2$ is energetically favorable to form from the elements.

\subsection{Electronic Properties}
The calculated electronic band structure and projected density of states (PDOS) for monolayer SnTe$_2$ are presented in Figure \ref{Fig1}c. The band structure clearly shows an overlap of the valence and conduction bands at the Fermi level (set to 0.52 eV), which is a definitive signature of metallic character. This finding is consistent with the 1T crystal phase, which is typically metallic for TMDs.~\cite{marinov2023ex,capobianco2022terahertz,fang2018structure,calandra2013,mortazavi2018,wen2015}

The partial density of states analysis reveals the orbital contributions to the electronic states. The states at energies below approximately -5.1 eV are dominated by \textit{s}-orbitals, while states at higher energies are primarily of \textit{p}-orbital character, reflecting the higher binding energy of \textit{s}-orbitals. Near the Fermi level, there is significant hybridization between the Te(\textit{p}) and Sn(\textit{s}) orbitals, which governs the material's conductive properties. A key distinction between SnTe$_2$ and conventional TMDs is the absence of \textit{d}-electron contributions near the Fermi surface. In transition metals, partially filled \textit{d}-orbitals form strong covalent bonds. Sn, being a post-transition metal, relies on weaker \textit{s--p} orbital bonding. This weaker bonding results in smaller cohesive energies and makes the material more susceptible to mechanical distortion, which is a desirable trait for flexible electronic applications. The higher density of states from Te(\textit{p}) orbitals compared to Sn(\textit{s}) orbitals near the Fermi level suggests that Te electrons are more delocalized and play a dominant role in electrical conduction.

\subsection{Vibrational Properties and Lattice Dynamics}
\begin{figure*}[htb!]
    \centering
    \includegraphics[trim = 0mm 0mm 0mm 0mm,width=\linewidth,clip=true]{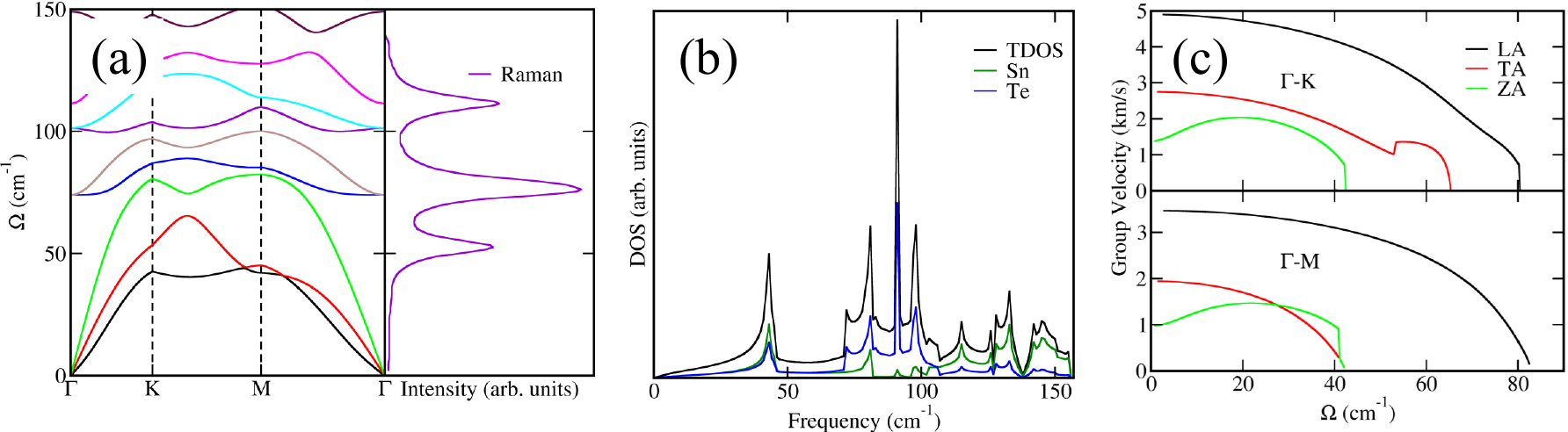}
    \caption{(a) Phonon dispersion spectrum of monolayer SnTe$_2$ together with the calculated spectroscopic intensities for Raman- and infrared-active modes. At the $\Gamma$ point, the vibrational modes decompose into the irreducible representation $\Gamma = A_{1g} + 2A_{2u} + 2E_{u} + E_{g}$. The $A_{1g}$ and $E_{g}$ modes are Raman active, while the $A_{2u}$ and $E_{u}$ modes are IR active, consistent with the symmetry of the $D_{3d}$ point group. (b) Total and partial phonon density of states (PDOS) for monolayer SnTe$_2$. The low-frequency region is dominated by Te vibrations, reflecting the heavier atomic mass, while Sn atoms contribute more prominently at higher frequencies. The overall soft spectrum, absence of a phonon bandgap, and strong overlap of Sn and Te contributions facilitate enhanced phonon-phonon scattering. (c) Phonon group velocities of monolayer SnTe$_2$ obtained by finite-difference evaluation of the slopes of the acoustic phonon branches near the $\Gamma$ point, shown along the (a) $\Gamma$--K and (b) $\Gamma$--M directions. The longitudinal acoustic (LA), transverse acoustic (TA), and flexural acoustic (ZA) modes all exhibit remarkably low velocities compared to typical two-dimensional crystals. The pronounced anisotropy between $\Gamma$--K and $\Gamma$--M further reflects bonding asymmetry, while the slow ZA mode underscores the vibrational softness that suppresses lattice thermal conductivity.}
    \label{Fig2}
\end{figure*}

The dynamical stability and vibrational characteristics of monolayer SnTe$_2$ were investigated by calculating its phonon dispersion spectrum, as shown in Figure \ref{Fig2}(a). The absence of any imaginary frequencies across the entire Brillouin zone confirms that the 1T structure is dynamically stable and corresponds to a true local minimum on the potential energy surface. From a thermoelectric standpoint, such dynamical stability is a prerequisite, since unstable modes would suppress phonon-limited transport.

The primitive unit cell contains three atoms, giving rise to nine phonon branches: three acoustic and six optical. Symmetry analysis yields Raman-active ($A_{1g}$, $E_g$) and infrared-active ($A_{2u}$, $E_u$) modes, consistent with the D$_{3d}$ point group.~\cite{keresztury2002raman} The Raman-active $E_g$ mode at 71.2 cm$^{-1}$ and the $A_{1g}$ mode at 107.4 cm$^{-1}$ are unusually soft compared to analogous modes in other related 2D materials such as SnS$_2$ (304.6 cm$^{-1}$) or SnSe$_2$ (185.7 cm$^{-1}$),~\cite{gurnani2018tin,fu2022controllable,herninda2024structural} where lighter chalcogen atoms stiffen the lattice. The softening of the vibrational frequencies in SnTe$_2$ arises primarily from the heavy mass of Te atoms, which lowers vibrational energies and stretches the acoustic branches. From a thermoelectric perspective, such soft, low--frequency phonons are desirable because they strongly suppress lattice heat conduction.~\cite{lin2017high,yuan2023soft} 

The acoustic branches (longitudinal acoustic, transverse acoustic, and the flexural ZA mode) are particularly revealing. Both the TA and ZA branches are very flat, especially near the zone center, indicating low phonon group velocities. Since the lattice thermal conductivity $\kappa_L$ scales roughly with the square of group velocity, these flat branches provide strong microscopic evidence for ultralow $\kappa_L$. In addition, the lack of a distinct phonon bandgap allows acoustic and optical branches to hybridize and interact more freely, increasing the density of allowed three-phonon processes. This expanded scattering phase space is known to enhance Umklapp phonon--phonon interactions, thereby accelerating heat dissipation, an effect widely observed in group-IV and transition-metal-based compounds and consistent with the ultralow thermal conductivity reported in prior studies.~\cite{zhang2017thermal,wang2019ultralow,zhang2020thermal} This combination of flat dispersions, heavy-element induced softening, and efficient scattering channels all converge toward reduced lattice heat transport, one of the central requirements for enhancing thermoelectric efficiency.

The phonon density of states in Figure \ref{Fig2}(b) further illustrates these points. The low-frequency region is dominated by Te vibrations, reflecting their heavy mass and strong influence on acoustic branches. Sn contributes more substantially at higher frequencies. This imbalance reduces the average phonon energy and strengthens phonon--phonon scattering, again limiting $\kappa_L$. Comparisons with other dichalcogenides highlight the unusual softness of SnTe$_2$: while SnS$_2$ and SnSe$_2$ exhibit Raman modes at higher frequencies, SnTe$_2$ exhibits its dominant vibrational features below 150 cm$^{-1}$. Such mode softening is an important thermoelectric asset, as it directly correlates with weaker heat conduction and more favorable conditions for high $ZT$.

\subsection{Thermal Transport and Optical Properties}

To quantify the thermal transport behavior more directly, we extracted the group velocities of the acoustic branches along the high-symmetry $\Gamma$–K and $\Gamma$–M paths, shown in Figure \ref{Fig2}(c). The group velocity $v_{g}$ of a phonon mode is formally defined as the first-order derivative of the phonon frequency $\omega(\mathbf{q})$ with respect to its wavevector $\mathbf{q}$: $v_{g}(\mathbf{q},s) = \nabla_{\mathbf{q}} \, \omega(\mathbf{q},s)$, where $\mathbf{q}$ is the phonon wavevector, $s$ is the phonon branch index (e.g., LA, TA, ZA), and $\omega(\mathbf{q},s)$ is the phonon dispersion relation. Along a one-dimensional path in reciprocal space (such as $\Gamma$–K or $\Gamma$–M), this reduces to a simple derivative, $v_{g}(\mathbf{q},s) = \frac{d \, \omega(\mathbf{q},s)}{d q}$, which we evaluated numerically by a finite-difference scheme. 

In practice, the slope of the dispersion curve near the long-wavelength (small $q$) limit gives the characteristic group velocity of the acoustic mode. In the long-wavelength limit, the LA and TA branches are nearly linear, while the ZA branch follows the quadratic dispersion expected for two-dimensional crystals. The calculated group velocities are remarkably low and anisotropic. Along $\Gamma$–K, the LA and TA modes reach maxima of 4980 m/s and 2900 m/s, respectively, while along $\Gamma$–M they fall to 3550 m/s and 2000 m/s. The flexural ZA mode is slower still, reinforcing the picture of weak phonon-mediated heat transport. These values are much smaller than graphene ($\sim 22,000$ m/s) and significantly lower than even common dichalcogenides such as MoS$_2$ (LA $\sim 11,000$ m/s).~\cite{Cai2013_MoS2_phonon,Wei2014_MLMoS2_vs_Graphene} The anisotropy between crystallographic directions further reflects bonding asymmetry, which limits coherent phonon propagation and contributes to the intrinsically low thermal conductivity.

To place these results in a broader transport framework, we recall that within the phonon Boltzmann transport formalism, the lattice thermal conductivity can be expressed as
$\kappa_{L} = \tfrac{1}{3} \sum_{s} C_{s}, v_{g}^{2}(s), \tau_{s}$, where $C_s$ is the mode-specific heat capacity, $v_g(s)$ is the phonon group velocity, and $\tau_s$ is the phonon lifetime of branch $s$. While both quantities formally contribute to $\kappa_L$, their physical roles are distinct. In monolayer SnTe$_2$, the exceptionally flat acoustic phonon dispersions impose intrinsically small group velocities, which already establish a strong, dispersion-driven suppression of lattice heat transport through the quadratic dependence on $v_g$. This mechanism is intrinsic to the lattice dynamics and does not rely on the explicit magnitude of phonon lifetimes. At the same time, qualitative signatures of reduced phonon lifetimes are implicitly evident in the calculated vibrational spectrum. In particular, the absence of a phonon bandgap and the strong overlap between acoustic and optical branches substantially increase the phase space for three-phonon scattering, including Umklapp processes, which are known to limit phonon lifetimes in soft, heavy-element systems. Thus, although phonon lifetimes are not computed explicitly in this work, their suppressive effect on thermal transport is encoded in the same vibrational features that give rise to the low group velocities. Taken together, these dispersion-controlled and scattering-enabled mechanisms provide a consistent microscopic explanation for the intrinsically low lattice thermal conductivity of monolayer SnTe$_2$.

The thermoelectric significance of these results is clear: low and anisotropic group velocities translate directly into low lattice thermal conductivity, since the group velocity enters explicitly into the lattice thermal conductivity expression, $\kappa_{L} = \frac{1}{3} \sum_{s} C_{s} v_{g}^{2}(s) \tau_{s}$, where $C_{s}$ is the mode-specific heat capacity, $v_{g}(s)$ is the group velocity, and $\tau_{s}$ is the phonon lifetime for branch $s$. The quadratic dependence on $v_{g}$ highlights that even modest reductions in phonon velocities can strongly suppress $\kappa_{L}$. For thermoelectric applications, the essential challenge is to decouple electrical and thermal transport. SnTe$_2$ appears to achieve this balance naturally. Its metallic electronic structure ensures adequate electrical conductivity, while its vibrational softness, evidenced by the remarkably low acoustic phonon velocities, ensures inefficient phonon-mediated heat transport. The heavy atomic mass of Te further suppresses phonon frequencies throughout the spectrum, and the absence of a phonon bandgap promotes Umklapp scattering, which reduces phonon lifetimes. Taken together, these features provide a consistent microscopic rationale for the intrinsically low thermal conductivity of SnTe$_2$, a property highly favorable for efficient thermoelectric performance.

\begin{figure}[htb!]
    \centering
    \includegraphics[trim = 0mm 0mm 0mm 0mm,width=\linewidth,clip=true]{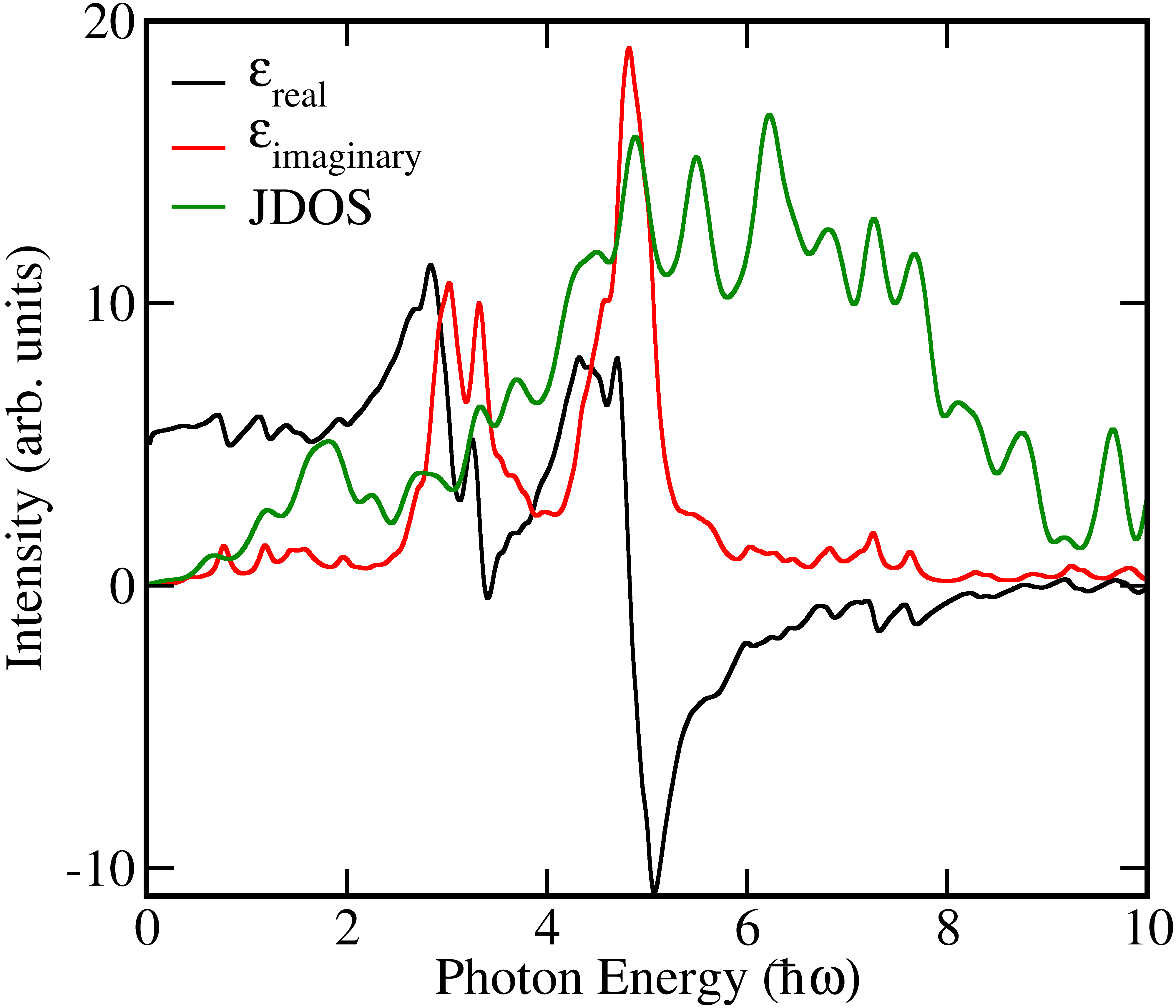}
    \caption{Calculated real ($\epsilon_{real}$) and imaginary ($\epsilon_{imaginary}$) parts of the dielectric function of monolayer SnTe$_2$ as a function of photon energy and the joint density of states (JDOS) indicating the density of possible interband electronic transitions as a function of photon energy.}
    \label{Fig3}
\end{figure}

The optical properties, although not directly governing lattice thermal transport, provide complementary insights into the energy conversion potential of SnTe$_2$.~\cite{zhang2025relationship} The dielectric function (Figure \ref{Fig3}) exhibits pronounced peaks at 3.02 and 3.35 eV, while the real part becomes negative around 4.84 eV, indicating the onset of plasmonic behavior. These features, corroborated by the joint density of states (Figure \ref{Fig3}), originate from interband transitions. Moreover, the relative intensities and sharpness of these peaks reflect the strong hybridization between Te \textit{p}-states and Sn \textit{p}-states identified in our electronic structure analysis, leading to allowed transitions with high oscillator strength. The absence of a well-defined optical bandgap in the low-energy region is consistent with the semimetallic character of SnTe$_2$, and the gradual rise in $\varepsilon(\omega)$ indicates sustained optical absorption across the visible and near-UV ranges. The plasmonic onset at 4.84 eV further corresponds to the collective oscillation of conduction electrons, which is expected for systems with substantial free-carrier density, as also supported by the finite density of states at the Fermi level. From a thermoelectric perspective, such optical resonances are significant because they can modify carrier populations through photoexcitation and drive local thermal gradients via the photothermoelectric effect. In addition, plasmonic responses can enhance light--matter interaction, potentially coupling optical absorption to heat management and charge transport. Thus, while secondary to the intrinsic lattice dynamics, these optical features are also noteworthy when compared to other 2D chalcogenides, where peak energies in a similar range typically indicate strong interband coupling and enhanced optical conductivity. In SnTe$_2$, the magnitude of $\varepsilon(\omega)$ at the principal resonance suggests efficient photon absorption and rapid carrier excitation, which can influence energy transport under photoexcitation. The strong optical response of SnTe$_2$ establishes a direct link to photothermoelectric functionality, broadening its application space beyond conventional thermoelectrics into optically assisted or hybrid energy-conversion technologies.

\section{Conclusion}
In summary, we have carried out a comprehensive first-principles study of the structural, electronic, vibrational, thermal, and optical properties of monolayer 1T-SnTe$_2$. Our results establish that this post-transition metal dichalcogenide is both energetically and dynamically stable, with a reasonably high cohesive energy and a phonon spectrum free of imaginary modes. The metallic electronic structure, arising from strong \textit{s}--\textit{p} orbital hybridization near the Fermi level, distinguishes SnTe$_2$ from conventional $d$-orbital TMDs and suggests inherent mechanical flexibility, which is advantageous for integration into flexible electronic platforms. The vibrational and thermal transport analyses highlight features that are directly beneficial for thermoelectric applications. The low phonon frequencies induced by the heavy tellurium atoms, combined with flat acoustic branches and small group velocities, point to intrinsically low lattice thermal conductivity. The absence of a phonon bandgap further facilitates phonon–phonon scattering, which is expected to enhance thermal resistance. Together, these lattice dynamical characteristics provide a microscopic rationale for considering monolayer SnTe$_2$ as a promising thermoelectric candidate where efficient heat suppression is required alongside metallic electrical conduction. The calculated Raman- and infrared-active modes offer clear spectroscopic fingerprints for experimental verification, while the optical response, dominated by interband transitions and showing a plasmonic resonance near 4.84 eV, indicates potential opportunities for photothermoelectric or optoelectronic applications. Altogether, this study identifies monolayer SnTe$_2$ as a 2D system that combines metallic conductivity with ultralow lattice thermal transport, two features rarely coexisting in 2D materials.~\cite{zhang2017thermal} These findings provide a strong theoretical foundation for experimental exploration of SnTe$_2$, and more broadly, open a pathway toward leveraging post-transition metal dichalcogenides in next-generation thermoelectric and energy conversion technologies.

\vspace{0.2cm}
\noindent \textbf{Funding Declaration}

\noindent This work was carried out as part of the Carnegie Mellon Diaspora Program. The authors gratefully acknowledge the use of computational resources provided by the Lehigh University High-Performance Computing infrastructure. Work at Lehigh University is supported in part by the U.S. Department of Energy, Office of Science, Basic Energy Sciences under Award DOE-SC0024099.

\vspace{0.2cm}
\noindent \textbf{Declaration of Competing Interest}

\noindent The authors declare that they have no known competing financial interests or personal relationships that could have appeared to influence the work reported in this paper.

\vspace{0.2cm}
\noindent \textbf{Authorship Contribution Statement}

\vspace{0.2cm}
\noindent \textbf{Author Contributions:} 
\textbf{K. A.:} Data Curation, Methodology, Visualization, Validation, Formal Analysis, Writing-Original Draft, Writing-Review \& Editing.
\textbf{J. E. E.:} Visualization, Validation, Formal Analysis, Writing-Original Draft, Writing-Review \& Editing.
\textbf{A. O. O.:} Data Curation, Methodology, Visualization, Validation, Formal Analysis, Writing-Original Draft, Writing-Review \& Editing.
\textbf{S. O. E.:} Visualization, Validation, Formal Analysis, Writing-Original Draft, Writing-Review \& Editing.
\textbf{S. M. M.:} Visualization, Validation, Formal Analysis, Writing-Original Draft, Writing-Review \& Editing.
\textbf{S. A.:} Visualization, Validation, Formal Analysis, Writing-Original Draft, Writing-Review \& Editing.
\textbf{M. N. A.:} Visualization, Validation, Formal Analysis, Writing-Original Draft, Writing-Review \& Editing.
\textbf{R. A. M.:} Visualization, Validation, Formal Analysis, Writing-Original Draft, Writing-Review \& Editing.
\textbf{A. C. I.:} Visualization, Validation, Formal Analysis, Writing-Review \& Editing. 
\textbf{F. I. E.:} Visualization, Validation, Formal Analysis, Writing-Review \& Editing.
\textbf{C. E. E.:} Conceptualization, Validation, Writing-Review \& Editing, Resources, Supervision, Funding Acquisition.

\vspace{0.2cm}
\noindent \textbf{Data Availability} 

\noindent The data supporting the findings of this study are included in the article. Additional data are available from the corresponding author upon reasonable request.

\bibliographystyle{elsarticle-num}

\providecommand{\noopsort}[1]{}\providecommand{\singleletter}[1]{#1}%

\end{document}